\documentclass[epj]{svjour}
\usepackage{graphics}
\begin{document}
\title{Quantum Computation in a radio-single mode cavity: the dissipative Jaynes and Cummings Model}

\author{S.Nicolosi\inst{1} \and P.Ferrante\inst{2}\and G.Scaccianoce\inst{2}\and G.Rizzo\inst{2}}

\institute{CNR-LATO (Laboratorio di Tecnologie Oncologiche), Cefal\'{u}, Italy \and DEIM (Dipartimento di Energia,
Ingegneria dell'Informazione e Modelli Matematici), Universit\'{a}
degli Studi di Palermo, Italy}

\date{Received: date / Revised version: date}
\abstract{
In this paper we have considered the interaction of a Jaynes and
Cummings system with the electromagnetic field in its vacuum state and,
solving the dynamical problem, we have analyzed the amount of
entanglement induced in the bipartite system (atom + cavity mode) by the common electromagnetic reservoir. This has allowed us to quantitatively
characterize the regime under which field-induced cooperative
effects are not vanished by dissipation. Once the Decoherence Free
Regime is reached, transient entanglement tends to become stationary
and, therefore, usable for quantum gate implementation.
\PACS{
      {03.65.Yz}{Decoherence, open systems, quantum statistical methods}   \and
      {03.67.Mn}{Entanglement production
and manipulation}{42.50.Fx}{Cooperative phenomena in quantum optical systems}
     } % end of PACS codes
} %end of abstract
\maketitle
\section{Introduction}
\label{intro}
The idea of implementing quantum information devices based on the
use of single atoms or molecules has gone progressively growing up
in the course of the last few years. The reason can be envied in the
high contemporary evolution of theoretical dynamical models along
with the ability reached by experimentalists in manipulating quantum
objects first considered {\it theoretician's tools} \cite{raimond}.
The state of art is that, although it has been possible to obtain
entanglement conditions between elementary systems in different
physical scenarios \cite{raimond}, the temporal persistence of
quantum coherence is an open problem. In this moment, it is
therefore necessary to put attention on the theoretical aspects of
decoherence in order to predict the experimental conditions under
which it is maximally reduced \cite{palma}. Almost Decoherence Free
Substates seem to have the characteristics of eligibility needed to
implement quantum computation. Their generation and temporal
persistence can be theoretically predicted in some high symmetrical
models
\cite{nud,Nicolosi,NicolosiPRL,tesi,Plenio1999,Pellizzari,Ficek03}.
The application of the formal solution of the
Markovian Master Equation (Nud theorem) \cite{nud} has supplied, in
the few analyzed systems, the prediction of a conditional building
up of entanglement. The result is obtained under symmetrical
condition corresponding to specific locations of involved subsystems
(atoms, molecules). Unfortunately, despite the positive results in
isolating single quantum objects \cite{raimond}, the difficulties
connected with the location of more than one atoms in fixed
arbitrary points is an open task.

Contrarily, the isolation of
a single atom for enough long time inside a cavity is today
possible. A single mode cavity and a two level
atom may be considered a hybrid two qubits system. Moreover, the Jaynes and Cummings Model describing this system is one of the few
exactly soluble models in quantum optics. It predicts many
interesting non classical states experimentally testable in
laboratory \cite{raimond}. In realistic situation, however, one
performs experiments in cavity with finite Q and in presence of
atomic spontaneous emission. So it become of fundamental importance
to know how the predictions of this model are affected by the
unavoidable presence of loss mechanisms. This problem is currently
very extensively studied, the main approximation being the
assumption of two different reservoirs, one for the atom and one for
the cavity mode respectively \cite{raimond2}. The Lindblad
master equation describing this model is easy to solve, the solution being the total destruction of
coherences because of the two different and nonspeaking channels of
dissipation. Contrarily to this approach, we assume a common bath of interaction between the cavity mode and the atom. The master equation so derived contains the simple one as
particular case. In this paper we show to be able to find the exact
solution of a dissipative J-C model assuming a common reservoir for
the bipartite system, which, on the ground of the above
consideration, appears to be a more realistic hypothesis. This leads
to the prediction of new cooperative effects, induced by the
zero-point fluctuations of environment, between the atom and the
cavity mode as the creation of conditional transient entanglement,
tending to become stationary as the strengths of the coupling with
the reservoir take a well defined value. Finally, in order to be
maximally realistic, we consider also
the loose of energy due to the imperfect reflection on the mirrors.
This correction in the microscopic model does not introduce
complication in the solution of the relative master equation because
the new bath is independent from the first one and, indeed, easily
treating from a theoretical point of view (not induced coherences).
In presence of the second channel of decoherence the building up of
entanglement exists during the transient period in which the atom is
confined inside the cavity. The long time solution (the order of
magnitude of the time involved is given in the next sections) shows
that the introduction of the second bath makes disappear coherence
(Rabi oscillation) among the two subsystem involved: the atom and
the cavity mode. Despite this fact, the time involved in decoherence
process can be made much longer than those necessary to implement a
quantum protocol as deducible from the theoretical analysis here
developed if interfaced with the experimental measures performed by
Haroches' group. The measure of the probability to find the atom in the excited state appear well fitted by the theoretical model here proposed. The standard models (two different baths) are
able to reproduce only the top of the curve of atomic population. Instead
our is able to reproduce also the lower part.

The paper is structured as follow: in section II we report the
principal step and approximation leading to the microscopic
derivation of the Markovian Master Equation and we solve it
when $T=0$, showing also the full equivalence between MME and
Piecewise Deterministic Processes (PDP) \cite{nud}. In section III
we apply the NuD theorem to derive the solution to the dissipative
J-C model. In section IV we report the main conclusion of the paper.
\section{The Markovian Master Equation}
\label{sec:1}
It is well know that under the Rotating Wave and the Born-Markov
approximations the master equation describing the reduced dynamical
behavior of a generic quantum system linearly coupled to an
environment can be put in the form \cite{nud,F.Petruccione}
\begin{equation}\label{ME}
\dot{\rho_S} (t)=-i[H_S+H_{LS},\rho_S (t)]+D(\rho_S (t)),
\end{equation}
where $H_S$ is the Hamiltonian describing the free evolution of the
isolated system,
\begin{eqnarray}\label{dissME}
\nonumber D(\rho_S (t))&=& \sum_{\omega } \sum_{\alpha ,\beta}
 \gamma_{\alpha ,\beta} (\omega)
(A_{\beta} (\omega ) \rho_S (t) A^{\dag}_{\alpha} (\omega
)\\ &-& \frac{1}{2}\{A^{\dag}_{\alpha} (\omega ) A_{\beta} (\omega
), \rho_S (t)\}),
\end{eqnarray}

\begin{equation}\label{lambshME}
 H_{LS} =\sum_{\omega}\sum_{\alpha,\beta}S_{\alpha ,\beta}
 (\omega)A^\dag_{\alpha}(\omega)A_{\beta}(\omega),
\end{equation}

\begin{equation}
 S_{\alpha ,\beta} (\omega)=\frac{1}{2i}(\Gamma_{\alpha ,\beta}
 (\omega)-\Gamma^\ast_{\beta ,\alpha}(\omega))
\end{equation}
and
\begin{equation}
 \gamma_{\alpha ,\beta} (\omega)=\Gamma_{\alpha ,\beta}
 (\omega)+\Gamma^\ast_{\beta ,\alpha}(\omega),
\end{equation}

$\Gamma_{\alpha ,\beta}(\omega)$ being the one-sided Fourier
transforms of the reservoir correlation functions. Finally we
recall that the operators $A_{\alpha}(\omega)$ and
$A_{\alpha}^{\dag}(\omega)$, we are going to define and whose
properties we are going to explore, act only in the Hilbert space
of the system.

Eq.\ (\ref{ME}) has been derived under the hypothesis that the
interaction Hamiltonian between the system and the reservoir, in the
Schr\"{o}dinger picture, is given by \cite{F.Petruccione}

\begin{equation}\label{HI}
H_I=\sum_{\alpha} A^{\alpha}\otimes B^{\alpha},
\end{equation}
that is the most general form of the interaction.

In the above expression $A^{\alpha}=(A^{\alpha})^\dag$ and
$B^{\alpha}=(B^{\alpha})^\dag$ are operators acting respectively on
the Hilbert space of the system  and  of the reservoir. The eq.\
(\ref{HI}) can be written in a slightly different form if one
decomposes the interaction Hamiltonian into eigenoperators of the
system and reservoir free Hamiltonian.

\begin{definition}{Definition}
Supposing the spectrum of $H_S$ and $H_B$ to be discrete
(generalization to the continuous case is trivial) let us denote the
eigenvalue of $H_S$ ($H_B$) by $\varepsilon$ ($\eta$) and the
projection operator onto the eigenspace belonging to the eigenvalue
$\varepsilon$ ($\eta$) by $\Pi (\varepsilon)$ ($\Pi (\eta)$). Then
we can define the operators:
\begin{equation}\label{AdiOmega}
A_{\alpha} (\omega)\equiv\sum_{\varepsilon^{'}
-\varepsilon=\omega} \Pi (\varepsilon) A_{\alpha}\Pi
(\varepsilon^{'}),
\end{equation}
\begin{equation}\label{AdiOmega}
B_{\alpha} (\omega)\equiv\sum_{\eta^{'} -\eta=\omega} \Pi (\eta)
B_{\alpha}\Pi (\eta^{'}).
\end{equation}
\end{definition}

From the above definition we immediately deduce the following
relations
\begin{eqnarray}\label{com1}
[H_S, A_{\alpha} (\omega)]=-\omega A_{\alpha} (\omega), \\
\nonumber [H_B, B_{\alpha} (\omega)]=-\omega B_{\alpha} (\omega),
\end{eqnarray}

\begin{eqnarray}\label{com2}
[H_S, A^{\dag}_{\alpha} (\omega)] &=& +\omega A^{\dag}_{\alpha}
(\omega)\;\;\; and \\
\nonumber [H_B, B^{\dag}_{\alpha}
(\omega)] &=& +\omega B^{\dag}_{\alpha} (\omega).
\end{eqnarray}
An immediate consequence is that the operators $A^{\dag}_{\alpha}
(\omega>0)$ e $A_{\alpha} (\omega>0)$ raise and lower the energy
of the system $S$ by the amount $\hbar \omega$ respectively and
that the corresponding interaction picture operators take the form

\begin{eqnarray}
e^{iH_{S} t}A_{\alpha} (\omega) e^{-iH_{S} t}=e^{-i\omega t}
A_{\alpha} (\omega), \\
\nonumber e^{iH_{B} t}B_{\alpha} (\omega)
e^{-iH_{B} t}=e^{-i\omega t} B_{\alpha} (\omega),
\end{eqnarray}

\begin{eqnarray}
e^{iH_{S} t}A^{\dag}_{\alpha} (\omega) e^{-iH_{S} t}=e^{+i\omega
t} A^{\dag}_{\alpha} (\omega) \;\;\; and \\
\nonumber e^{iH_{B}
t}B^{\dag}_{\alpha} (\omega) B^{-iH_{B} t}=e^{+i\omega t}
B^{\dag}_{\alpha} (\omega).
\end{eqnarray}

Finally we note that
\begin{eqnarray}\label{AconAlfadiOmega}
A^{\dag}_{\alpha} (\omega)=A_{\alpha} (-\omega) \;\;\; and \;\;\;
B^{\dag}_{\alpha} (\omega)=B_{\alpha} (-\omega).
\end{eqnarray}
Summing eq.\ (\ref{AconAlfadiOmega}) over all energy differences
and employing the completeness relation we get

\begin{eqnarray}
\sum_{\omega}A^{\dag}_{\alpha} (\omega)=\sum_{\omega}A_{\alpha}
(-\omega)=A_\alpha  \;\;\; and \\
\nonumber \sum_{\omega}B^{\dag}_{\alpha}
(\omega)=\sum_{\omega}B_{\alpha} (-\omega)=B_\alpha
\end{eqnarray}
The above positions enable us to cast the interaction Hamiltonian
into the following form
\begin{eqnarray}
H_I=\sum_{\alpha ,\omega,\omega '} A_{\alpha} (\omega) \otimes
B_{\alpha}(\omega ')= \\
\nonumber \sum_{\alpha ,\omega,\omega '}
A^{\dag}_{\alpha} (\omega) \otimes B^{\dag}_{\alpha}(\omega ').
\end{eqnarray}
The reason for introducing the eigenoperator decomposition, by
virtue of which  the interaction Hamiltonian in the interaction
picture can now be written as
\begin{equation}\label{intgen}
H_I(t)=\sum_{\alpha ,\omega,\omega '}e^{-i(\omega+\omega ')t}
A_{\alpha} (\omega) \otimes B_{\alpha}(\omega '),
\end{equation}
is that exploiting the rotating wave approximation, whose
microscopic effect is to drop the terms for which $\omega\neq
-\omega'$, is equivalent to the Schrodinger picture interaction
Hamiltonian:
\begin{equation}
H_I=\sum_{\alpha ,\omega} A_{\alpha} (\omega) \otimes
B_{\alpha}(-\omega)=\sum_{\alpha ,\omega} A_{\alpha} (\omega)
\otimes B^{\dag}_{\alpha}(\omega).
\end{equation}

\begin{theorem}{Lemma} \label{TH1}
The Rotating Wave Approximation imply the conservation of the free
energy of the global system, that is
\begin{equation}
 [H_S+H_B,H]=0
\end{equation}

\end{theorem}

\subsection{Proof}
\label{sec:2}
The necessary condition involved in the previous proposition is
equivalent to the equation $[H_S+H_B,H_I]=0$ we are going to
demonstrate.

\begin{eqnarray}\label{cons}
[H_S+H_B,H] &=& [H_S+H_B,H_I] \\ \nonumber &=& [H_S,H_I]+[H_B,H_I] \\ \nonumber
&=& \sum_{\alpha ,\omega } [H_S,A_{\alpha} (\omega)] \otimes
B_{\alpha}^{\dag}(\omega) + \\
\nonumber  && \sum_{\alpha ,\omega } A_{\alpha}
(\omega) \otimes [H_B,B_{\alpha}^{\dag}(\omega)] \\ \nonumber
&=&-\sum_{\alpha ,\omega}\omega A_{\alpha} (\omega) \otimes
B_{\alpha}(-\omega)+ \\
\nonumber && \sum_{\alpha ,\omega}\omega A_{\alpha}
(\omega) \otimes B_{\alpha}(-\omega)=0.
\end{eqnarray}

where we have made use of eq. (\ref{com1},\ref{com2})
\rule{5pt}{5pt}

\begin{theorem}{Lemma} \label{TH2}
The detailed balance condition in the thermodynamic limit imply
\cite{Alicki}
\begin{equation}
 \gamma_{\alpha \beta}(\omega)=e^{-\beta \omega} \gamma_{\alpha
 \beta}(-\omega)
\end{equation}
\end{theorem}
where $\beta=(k_B T)^{-1}$ \rule{5pt}{5pt}
\begin{theorem}{Corollary} \label{TH3}
Let us suppose the temperature of the thermal reservoir to be the
absolute zero, on the ground of Lemma 2 immediately we see that
\begin{equation}
 \gamma_{\alpha \beta}(\omega<0)=0\;\;\;\rule{5pt}{5pt}
\end{equation}
\end{theorem}

Let us now  cast eq.\ (\ref{ME}) in a slightly different form
splitting the sum over the frequency, appearing in eq.\
(\ref{dissME}), in a sum over the positive frequencies and a sum
over the negative ones so to obtain

\begin{eqnarray}\label{Diss}
&&\nonumber D(\rho_S (t))\\ \nonumber &=&\sum_{\omega>0 ,\alpha
,\beta} \gamma^{\alpha ,\beta} (\omega) (A^{\beta} (\omega) \rho_S
A^{\alpha \dag} (\omega)\\\nonumber &-&\frac{1}{2}\{A^{\beta \dag}
(\omega) A^{\alpha} (\omega),\rho_S \})\\\nonumber &+&\sum_{\omega
>0 ,\alpha ,\beta} \gamma^{\alpha ,\beta} (-\omega)
(A^{\alpha \dag} (\omega) \rho_S A^{\beta} (\omega)
\\&-&\frac{1}{2}\{A^{\alpha} (\omega) A^{\beta \dag} (\omega),\rho_S
\}),
\end{eqnarray}
where we again make use of eq.\ (\ref{AconAlfadiOmega}). In the
above expression we can recognize the first term as responsible of
spontaneous and stimulated emission processes, while the second one
takes into account stimulated absorption, as imposed by the lowering
and raising properties of $A^{\alpha}(\omega)$. Therefore if the
reservoir is a thermal bath at $T=0$ the corollary 4 tell us that
the correct dissipator of the Master Equation can be obtained by
suppressing the stimulated absorption processes in eq.\
(\ref{Diss}).
\subsection{NuD Theorem}
\label{sec:3}
We are now able to solve the markovian master equation when the
reservoir is in a thermal equilibrium state characterized by $T=0$.
We will solve a Cauchy problem assuming the factorized initial
condition to be an eigenoperator of the free energy $H_S+H_B$. This
hypothesis doesn't condition the generality of the found solution
being able to extend itself to an arbitrary initial condition
because of the linearity of the markovian master equation \footnote{
It is out of relevance to consider initial condition having non-zero
coherence between the environment and the system because it is not
possible to resolve them in the reduced dynamics obtained tracing on
the environment degrees of freedom.}.

\begin{theorem}{NuD theorem} \label{TH4}
If eq. (\ref{ME}) is the markovian master equation describing the
dynamical evolution of a open quantum system S, coupled to an
environment B, assumed to be in the detailed-balance thermal
equilibrium state characterized by a temperature T=0, and if the
global system is initially prepared in a state $\rho(0)=\rho_B(0)
\rho_S(0)$ so that $(H_S+H_B)\rho(0)(H_S+H_B)=E_L^2 \rho(0)$, where
$E_L=E_S+E_B$ is the free energy of the global system then
$\rho_S(t)$ is in the form of a Piecewise Deterministic Process
\cite{F.Petruccione}, that is a process obtained combining a
deterministic time-evolution with a jump process.
\end{theorem}

The proof of the theorem is contained in the paper \cite{nud}. My
aim here is to give an explanation of the found implication.

A PDP is a statistical mixture of alternative generalized
trajectories evolving individually in a deterministic way. This
statement is mathematically given by the equation
\begin{eqnarray}\label{somma}
\rho_S(t)=\sum_{i=0}^N\rho_{i}(t),
\end{eqnarray}

where the quantum trajectories $\rho_{i}(t)$ are obtained by the deterministic non-unitary equation

\begin{equation}
\rho_i (t)=U(t)f_i (t) U^{\dag} (t),
\end{equation}
where, in particular,
\begin{equation}
f_{N} (t)=\rho_{N} (0)
\end{equation}
and $U(t)=e^{-\frac{i}{\hbar}Bt}$, $U^\dag
(t)=e^{\frac{i}{\hbar}B^\dag t}$, $B$ being
\begin{equation}
 B=H_0-\frac{i}{2\hbar}\sum_{\omega>0 ,\alpha ,\beta}
\gamma^{\alpha \beta} (\omega)A^{\beta
\dag}(\omega)A^{\alpha}(\omega)\equiv H_0-\frac{i}{2\hbar}H',
\end{equation}
with $H'$ hermitian. Finally,
\begin{eqnarray}\label{brutta}
&f_{N-j}& (t)=\sum_{\omega ',\alpha ',\beta '}\sum_{\omega
",\alpha ",\beta "} \\ \nonumber &...& \sum_{\omega^j ,\alpha^j
,\beta^j}\gamma^{\alpha \beta} (\omega)\gamma^{\alpha ' \beta '}
(\omega ')\gamma^{\alpha "\beta "} (\omega ")...\gamma^{\alpha^j
\beta^j} (\omega^j) \\ \nonumber &\times & \int_0^t
\int_0^{t'}\int_0^{t"}...\int_0^{t^j}dt'dt"...dt^j \\ \nonumber &.&
U^{-1}(t')A^{\beta '}(\omega ')U(t')U^{-1}(t")A^{\beta "}(\omega
")U(t")\\ \nonumber &...&
U^{-1}(t^j)A^{\beta^j}(\omega^j)U(t^j)f_{N} (t^j)
U^{\dag}(t^j)A^{\alpha^j \dag}(\omega^j)U^{\dag -1}(t") \\
\nonumber &...& U^{\dag}(t")A^{\alpha " \dag}(\omega ")U^{\dag
-1}(t")U^{\dag}(t')A^{\alpha ' \dag}(\omega ')U^{\dag -1}(t'),\\ \nonumber &j& =1,...,N \;\;\; \rule{5pt}{5pt}
\end{eqnarray}

These last are {\it generalized} respect to F.Petruccione and
H.J.Carmochael approach, which leads to $\rho_S(t)=\sum_i
|\psi_i><\psi_i|$. The last expansion, in terms of proper
trajectories, is obtainable from ours if and only if we are able
to put into diagonal form the spectral correlation tensor, that is
known to be always possible because of the positivity of
$\gamma^{\alpha \beta}$, but nobody is able to do it, with exception
of few highly symmetrical systems.

The found solution (NuD theorem) ensures that the dynamical
processes, whose statistical mixture gives the open system
stochastic evolution, are deterministic. This demonstrates that the
evolution is representable as a Piecewise Deterministic Process
(PDP) \cite{F.Petruccione}. The found solution generalizes the PDPs
introduced by H.J.Carmichael and formalized by F.Petruccione and
H.P.Breuer. Actually, it is applicable also when the Markovian
Master Equation isn't in the Lindblad form. This, as already
highlighted, in general, introduces simplification in the further
calculations, but because of the difficulty to recast the equation
in this form the results obtained are in general merely formal.
Tough the eq. (\ref{brutta}) seems complicated to use it is a
powerful predictive tool. We have tested it deriving the
photocounting formula \cite{tesi,Davies}; reproducing the
environment-induced entanglement between two two-level
not-direct-interacting atoms placed in fixed arbitrary points in the
free space \cite{tesi,Plenio1999,Pellizzari,Ficek03} and Carmichael
unravelling of the Master Equation \cite{tesi,Carmichael}.

Moreover, we have tested the NuD theorem's predictive capability
solving the dynamics of two two-level dipole-dipole interacting
atoms placed in fixed arbitrary points inside a single mode cavity
in presence of atomic spontaneous emission and cavity losses
\cite{NicolosiPRL}; $n$ two-level not-direct-interacting atoms
placed in fixed arbitrary points inside a single mode cavity in
presence of atomic spontaneous emission and cavity losses
\cite{Nicolosi}; a bipartite hybrid model, known as Jaynes-Cummings
model, constituted by an atom and a single mode cavity linearly
coupled and spontaneously emitting in the same environment (next
subsection) and two harmonic oscillator linearly coupled and
spontaneously emitting in the same environment (work in progress).

\section{Dissipative Jaynes and Cummings model}
\label{sec:4}
The Jaynes-Cummings Model describes, under the Rotating Wave Approximation, the resonant
interaction between a single two-level atom and the single mode of
the electromagnetic field protected by a perfect cavity (no loosing
of energy). The model has been introduced in 1963 \cite{Jaynes} in
order to analyze the classical aspects of spontaneous emission and
to understand the effects of quantization on the atomic evolution.
Actually, despite its apparent simplicity, this model has revealed
interesting non-classical proprieties characterizing the
matter-radiation interaction. Moreover, thanks to the recent
experimental implementation of high Q cavities, it is today possible
to verify the most of the theoretical predictions of the model
\cite{raimond,raimond2,raimond3}.

The major experimental limitation is related to the coupling with a
chaotic environment able to destroy the quantum coherences. A
theoretical approach including the loss of energy due to the
interaction of the atom and the cavity mode with the free
electromagnetic field is more complete and, as we will show, it is
suitable to reproduce the experimental measured decay of the
population of the atom \cite{raimond}. In particular, assuming a
common bath of interaction between the cavity mode and the atom, the
theoretical probability to find the atom in the excited state
performs Rabi oscillation exponentially decaying. This fact is
consistent with the open dynamics but it is not the only effect.
Actually, the common bath induces cooperation between the two
involved parts (mode and atom). This behavior competes with the
exponential decay. In the long time limit the the exponential decay
wins on cooperation if we work under the experimental condition
performed by Haroche group. In the paper \cite{raimond} is reported
the experimental graph relative to the probability to find the atom
in its excited state as a function of the time. We can interpret the
upper part of the figure as the exponential decay and the lower part
of it (increasing of probability) as the cooperation induced by the
common reservoir. The new theoretical approach, here presented, is
better than the usual one (two independent baths) because it is able
to reproduce the experimental curve in a complete way. In fact the
two bath approximation keeps account only for the dissipation
meaning that the Rabi oscillation of the atomic population goes to
zero every period characterizing the free dynamics of the bipartite
system. In this case the cooperative part of the dynamics
disappears: the two parts do not speak trough the common bath, the
main behavior being the dissipation of energy in the reservoirs.
Moreover it is possible to demonstrate that single bath approach is
more general than the other including it as particular case. This
fact is very well understood if the parts are, for example, two or
more atoms, in which case the cooperation is the maximum one if the
distance among atoms is small enough and it reaches its minimum
value when the distance goes to infinity \cite{Nicolosi}. In the
last case the out diagonal terms of the spectral correlation tensor
go to zero meaning that the parts see independent reservoirs. The
lack of a microscopical derivation of the coupling constant of the
mode with the electromagnetic field makes difficult the analytical
derivation of an analogue relation in the case here analyzed.
Despite this fact it will be shown that the single bath case is more
general than the other one because the independent baths case does
not reproduce the out diagonal terms giving a simplified Master
Equation unable to reproduce part of the experimental measurements.

The Hamiltonian describing the open system is
\begin{eqnarray}
 H&=&\hbar\frac{\omega_0}{2}S_z+\hbar \omega_0 \alpha^\dag
 \alpha+\hbar(\epsilon\alpha S_+ +\epsilon^\ast\alpha^\dag S_-) \\ \nonumber &+&
\hbar\sum_{\vec{k},\lambda}\omega_{\vec{k}}
b^\dag_{\lambda}(\vec{k})b_{\lambda}(\vec{k})
  \\ \nonumber &+&\sum_{\vec{k},\lambda}[g^\ast_{\vec{k},\lambda} b_{\lambda}(\vec{k}) +g_{\vec{k},\lambda}
  b^{\dag}_{\lambda}(\vec{k})](S_+ +S_-) \\ \nonumber &+&
  \sum_{\vec{k},\lambda}[s^\ast_{\vec{k},\lambda}
  b_{\lambda}(\vec{k})+s_{\vec{k},\lambda}
  b^{\dag}_{\lambda}(\vec{k})](\alpha^\dag +\alpha).
\end{eqnarray}

If we make the position

\begin{eqnarray}
 H_S=\hbar\frac{\omega_0}{2}S_z+\hbar \omega_0 \alpha^\dag
 \alpha+\hbar(\epsilon\alpha S_+ +\epsilon^\ast\alpha^\dag S_-),
\end{eqnarray}

\begin{eqnarray}
 H_B=\hbar\sum_{\vec{k},\lambda}\omega_{\vec{k}}
b^\dag_{\lambda}(\vec{k})b_{\lambda}(\vec{k}),
\end{eqnarray}

\begin{eqnarray}
 H_I=\sum_{\alpha=1}^2 A^{\alpha}B^{\alpha},
\end{eqnarray}
where
\begin{eqnarray}
B^1=\sum_{\vec{k},\lambda}[g^\ast_{\vec{k},\lambda}
b_{\lambda}(\vec{k}) +g_{\vec{k},\lambda}
  b{^\dag}_{\lambda}(\vec{k})],
\end{eqnarray}

\begin{eqnarray}
B^2=\sum_{\vec{k},\lambda}[s^\ast_{\vec{k},\lambda}
b_{\lambda}(\vec{k})+s_{\vec{k},\lambda}
  b{^\dag}_{\lambda}(\vec{k})]
\end{eqnarray}
\begin{eqnarray}
A^1=(S_+ +S_-),
\end{eqnarray}
\begin{eqnarray}
A^2=(\alpha^\dag +\alpha)
\end{eqnarray}

we can describe the reduced dynamics of the bipartite system at $T=0$ by a Master Equation of the standard form
\begin{eqnarray}\label{jcme}
\dot{\rho_S} (t)=-\frac{i}{\hbar}[H_S,\rho_S (t)]+D(\rho_S (t)).
\end{eqnarray}

In this expression we have neglected the Lamb-Shift. This approximation is made possible because we have considered, ab initio, a direct linear static interaction among the parts respect to which the Lamb-Shift is negligible.  In the above equation

\begin{eqnarray}
 D(\rho_S (t))&=&\gamma^{1,1}(S_- \rho_S S_+ -\frac{1}{2}
\{S_+ S_-,\rho_S \}) \\ \nonumber &+&\gamma^{2,2}(\alpha\rho_S
\alpha^\dag -\frac{1}{2} \{\alpha^\dag \alpha ,\rho_S \}) \\
\nonumber &+&\gamma^{1,2}(\alpha \rho_S S_+ -\frac{1}{2}
\{\alpha^\dag S_-,\rho_S \}) \\ \nonumber &+&\gamma^{2,1}(S_-
\rho_S \alpha^\dag -\frac{1}{2} \{S_+ \alpha,\rho_S \})
\end{eqnarray}

where

\begin{eqnarray}
\gamma^{1,1} =\pi\sum_{\vec{k},\lambda }\mid g_{\vec{k}\lambda
}\mid^2
 \delta (\omega_{\vec{k}}-\omega )
\end{eqnarray}
\begin{eqnarray}
\gamma^{2,2} =\pi\sum_{\vec{k},\lambda }\mid s_{\vec{k}\lambda
}\mid^2
 \delta (\omega_{\vec{k}}-\omega )
\end{eqnarray}
\begin{eqnarray}
\gamma^{1,2} &=& \pi\sum_{\vec{k},\lambda } Re (
s^\ast_{\vec{k}\lambda} g_{\vec{k}\lambda })
 \delta (\omega_{\vec{k}}-\omega ) \\ \nonumber &+& i\hbar\pi\sum_{\vec{k},\lambda } Im (
s^\ast_{\vec{k}\lambda} g_{\vec{k}\lambda })
 \delta (\omega_{\vec{k}}-\omega )
\end{eqnarray}
\begin{eqnarray}
\gamma^{2,1} &=& \pi\sum_{\vec{k},\lambda } Re (
s^\ast_{\vec{k}\lambda} g_{\vec{k}\lambda })
 \delta (\omega_{\vec{k}}-\omega ) \\ \nonumber &-& i\hbar\pi\sum_{\vec{k},\lambda } Im (
s^\ast_{\vec{k}\lambda} g_{\vec{k}\lambda })
 \delta (\omega_{\vec{k}}-\omega )
\end{eqnarray}

The master equation for $\rho_S$ can be solved applying the NuD theorem to this case:

\begin{equation}
\rho_S (t)=\sum_{i=0}^{n} \rho_i (t),\; \; \; \; i\in N
\end{equation}

where $n$ is the number of the excitation initially given to the
system ($\rho_S (0)\equiv \rho_{n} (0)$) and $i$ is the index giving
the number of excitations characterizing every quantum trajectory.

The trajectories evolve in time in accordance with

\begin{equation}
\rho_i (t)=N(t)f_i (t) N^{\dag} (t),
\end{equation}

where $f_i(t)$ is given by eq.(\ref{brutta}) and $N(t)=e^{-\frac{i}{\hbar}}Bt$ is a nonunitary
temporal evolution operator, $B$ being, in general, non-hermitian as it appears from the following equation:

\begin{equation}
B=\frac{\hbar}{2}\Omega_0 S_z+\hbar\Omega\alpha^\dag\alpha+M S_+\alpha+P  \alpha^\dag S_- -i\frac{\gamma^{1,1}}{2},
\end{equation}

where
\begin{equation}
 \Omega_0=\omega_0-i\gamma^{1,1}
\end{equation}
\begin{equation}
 \Omega=\omega_0-i\gamma^{2,2}
\end{equation}
\begin{equation}
 M=\varepsilon-i\gamma^{2,1}
\end{equation}
\begin{equation}
 P=\varepsilon^{\ast}-i(\gamma^{2,1})^{\ast}.
\end{equation}

Let us suppose the system in the initial state characterized by $(n-1)$ excitations in the cavity mode and the atom in its excited state $\mid +>$:
\begin{equation}
\rho_S(0)=\mid (n-1),+><(n-1),+ \mid,
\end{equation}
then every quantum trajectory $\rho_i(t)$ belonging to statistical mixture characterizing the dynamical evolution of the system will have the form
\begin{eqnarray}
\rho_i(t)&=&\rho_{1,1_i}(t)\mid (i-1),+><(i-1),+ \mid \\ \nonumber &+& \rho_{1,2_i}(t)\mid (i-1),+><i,- \mid \\\nonumber &+& \rho_{2,1_i}(t)\mid i,-><(i-1),+ \mid \\ \nonumber &+& \rho_{2,2_i}(t)\mid i,-><i,- \mid.
\end{eqnarray}

The highest energy subspace ($i=n$) is easily solved and the block vector relative to this subspace has the form:
\begin{eqnarray}
\rho_{1,1_n}(t)&=&\frac{e^{-2\gamma_{2,2}\left(n-\frac{1}{2}\right)t-\gamma_{1,1}t}}{2\mid A_n\mid^2}\{(\mid A_n\mid^2-\mid \Delta\mid^2) \\ \nonumber &.& \cos(2a_nt)+(\mid A_n\mid^2+\mid \Delta\mid^2)\cosh(2b_nt) \\ \nonumber &-& 2ib_n\Delta(\sin(2a_nt)+\sinh(2b_nt))\}
\end{eqnarray}
\begin{eqnarray}
\rho_{1,2_n}(t)&=&\frac{i\sqrt{n}M^{\ast}}{2 A_n^{\ast}}e^{-2\gamma_{2,2}\left(n-\frac{1}{2}\right)t-\gamma_{1,1}t} \\ \nonumber &.& \{\sin(2a_nt)-i\sinh(2b_nt)\\\nonumber&+&i\frac{\Delta}{A_n}(\cos(2a_nt)-\cosh(2b_nt))\}
\end{eqnarray}
\begin{eqnarray}
\rho_{2,1_n}(t)=\rho^{\ast}_{1,2_n}(t)
\end{eqnarray}
\begin{eqnarray}
\rho_{2,2_n}(t)&=&\frac{n\mid M\mid^2}{4 \mid
A_n\mid^2}e^{-2\gamma_{2,2}\left(n-\frac{1}{2}\right)t-\gamma_{1,1}t} \\ \nonumber &.& \{\sin(2a_nt)-i\sinh(2b_nt)\},
\end{eqnarray}
where
\begin{eqnarray}
A_n&=&a_n+ib_n=\sqrt{\Delta^2+nMP}\\\Delta&=&\frac{1}{2}(\gamma_{1,1}-\gamma_{2,2}).
\end{eqnarray}

Let us note that if we have started from the initial condition $\mid
\psi_1 (0)>=\mid n,->$ we have obtained the same dynamical behavior.
This fact ensures that an arbitrary linear combination of the two
different initial condition will bring to the same dynamics. This
fact is really important because it is not simple to prepare one or
the other of the initial states. Actually, when we inject an
excitation inside the system we can only know that the system is in
a statistical mixture of the two states. But we have seen seen that
the dynamics is not case sensitive and therefore a statistical
mixture of the two states leads to the described dynamical
evolution.

\subsection{Entanglement building up}
\label{sec:5}
The circumstance that we succeed in finding the explicit time
dependence of the solution of the master equation (\ref{jcme})
provides an occasion to analyze in detail at least some aspects of
how entanglement is getting established in our bipartite system.
As particular case we can choose $n=1$ so obtaining in a simple
way the complete dynamics of the open system in the form
\begin{eqnarray}\label{statmix}
\rho_S(t)=\rho_0(t)+\rho_1(t),
\end{eqnarray}
where
\begin{eqnarray}
\rho_1(t)&=&\rho_{1,1_1}(t)\mid 0,+><0,+ \mid \\ \nonumber &+& \rho_{1,2_1}(t)\mid
0,+><1,- \mid \\\nonumber &+& \rho_{2,1_1}(t)\mid 1,-><0,+ \mid
\\ \nonumber &+& \rho_{2,2_1}(t)\mid 1,-><1,- \mid
\end{eqnarray}
and
\begin{eqnarray}
\rho_0(t)=(1-\rho_{1,1_1}(t)-\rho_{2,2_1}(t))\mid 0,-><0,- \mid
\end{eqnarray}
On the basis of the block diagonal form exhibited by Eq.
(\ref{statmix}), at a generic time instant $t$, the system is in a
statistical mixture of the vacuum state of the system and of a
one-excitation appropriate density matrix describing with certainty
the storage of the initial energy. In order to analyze the time
evolution of the degree of entanglement that gets established
between the two initially uncorrelated parties, we exploit the
concept of concurrence $C$ first introduced by Wootters
\cite{Wootters97,Wootters98}. If, at an assigned time $t$, no photon
have been emitted the conditional concurrence $C$ assumes the form:

\begin{eqnarray}
C(t)&=&\sqrt{\frac{2(\rho_{1,1_{n}}\rho_{2,2_{n}}+\mid\rho_{1,2_{n}}\mid^2)+4\rho_{2,2_{n}}\mid\rho_{1,2_{n}}}{\rho_{1,1_{n}}+\rho_{2,2_{n}}}}\\\nonumber&-&
\sqrt{\frac{2(\rho_{1,1_{n}}\rho_{2,2_{n}}+\mid\rho_{1,2_{n}}\mid^2)-4\rho_{2,2_{n}}\mid\rho_{1,2_{n}}}{\rho_{1,1_{n}}+\rho_{2,2_{n}}}}.
\end{eqnarray}

In the analyzed case ($n=1$), as clearly shown in Fig.1, obtained
using the experimental values setted by Haroche's group, the degree
of entanglement (C(t)) starting from zero increases during the transient collapsing to the initial value when time is long enough. This fact
depend on the choice of the atom whose spontaneous emission time
$\gamma_{1,1}$ is much longer than the cavity damping time
$\gamma_{2,2}$. In accordance to this fact the probability to find
the atom in the excited state starting from $1$ go to zero when
$t\gg (\gamma_{1,1}+\gamma_{2,2})^{-1}$ as clearly showed in Fig.2.
This Figure reproduce in a perfect way the experimental measures
performed by Haroche's group \cite{raimond}. The standard
theoretical models assume two different channel of dissipation (one
for the atom and one for the cavity mode). The corresponding Master
Equation is simpler to solve because of the absence of {\it
cooperative terms} \cite{nud} but the corresponding dynamics fits
only the upper part of the measures of the Haroche's group ({\it
dissipative behavior}). The low part of the graph represent the
cooperation induced by the common reservoir between the cavity mode
and the atom.

\begin{figure}
\resizebox{0.5\textwidth}{!}{
  \includegraphics{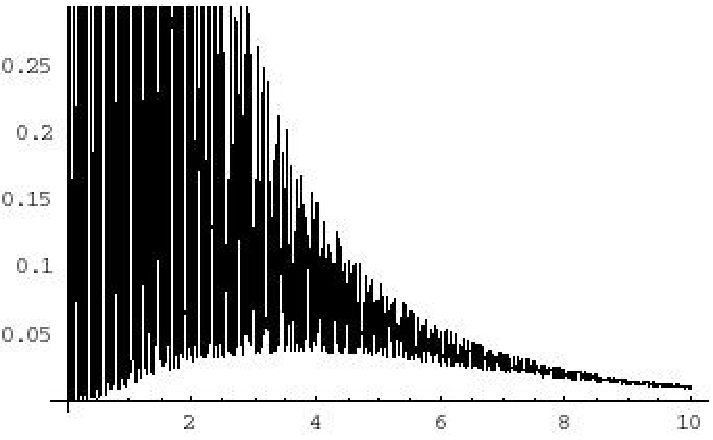}
}
%\vspace{5cm}       % Give the correct figure height in cm
\caption{$\gamma_{1,1}=1/30 KHz$, $\gamma_{2,2}=1 KHz$, $\gamma_{1,2}=\gamma_{2,1}=1/\sqrt{30} KHz$ , $\epsilon=47\pi KHz$
, t=10=1ms
}
\label{fig:1}       % Give a unique label
\end{figure}
\begin{figure}
\resizebox{0.5\textwidth}{!}{%
  \includegraphics{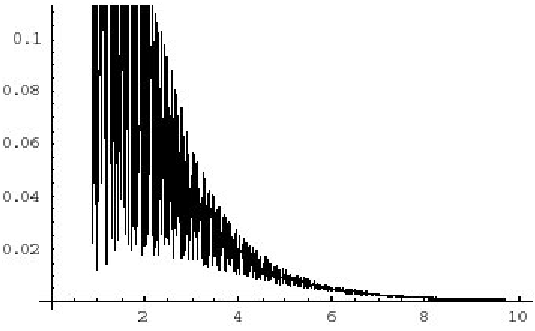}
}
%\vspace{5cm}       % Give the correct figure height in cm
\caption{$\gamma_{1,1}=1/30 KHz$, $\gamma_{2,2}=1 KHz$, $\gamma_{1,2}=\gamma_{2,1}=1/\sqrt{30} KHz$ , $\epsilon=47\pi KHz$
, t=10=1ms}
\label{fig:2}       % Give a unique label
\end{figure}

Such cooperation become the maximum one when
$\gamma_{1,1}=\gamma_{2,2}$, as clearly shown in Fig.3
and Fig.4. These ones depicted the concurrence and the probability
to find the atom in its excited state, respectively.

Under this condition (Decoherence Free Regime) Eq. (\ref{statmix})
suggests that, for $t\gg (\gamma_{1,1}+\gamma_{2,2})^{-1}$, the
correspondent asymptotic form assumed by $\rho_S(t)$ is time
independent and such that the probability of finding energy in the
bipartite system is $\frac{1}{2}$:
\begin{eqnarray}
\rho_S(t\gg (\gamma_{1,1}+\gamma_{2,2})^{-1})=\frac{1}{2}\mid
\psi_G><\psi_G\mid \\\nonumber+\frac{1}{2}\mid \psi_A><\psi_A\mid,
\end{eqnarray}
where
\begin{eqnarray}
\mid \psi_G>=\mid 0,-><0,-\mid
\end{eqnarray}
is the ground state of the bipartite system and
\begin{eqnarray}
\mid \psi_A>=\frac{1}{\sqrt{2}}(\mid 0,+><1,-\mid-\mid 1,-><0,+\mid)
\end{eqnarray}
is the maximally antisymmetric entangled state of the system.

This fact suggests that {\it stationary} entangled states of the JC
system can be generated by putting a single photon detector able to
capture in a continuous way all the excitations lost by the system
in the reservoir. Reading out the detector states at $t\gg
(\gamma_{1,1}+\gamma_{2,2})^{-1}$, if no photons have been emitted,
then, as a consequence of the measurement outcome, our system is
projected into the maximally antisymmetric entangled state
$\mid\psi>=\frac{1}{\sqrt{2}}(\mid 0,+><1,-\mid-\mid 1,-><0,+\mid)$.

\begin{figure}
\resizebox{0.5\textwidth}{!}{%
  \includegraphics{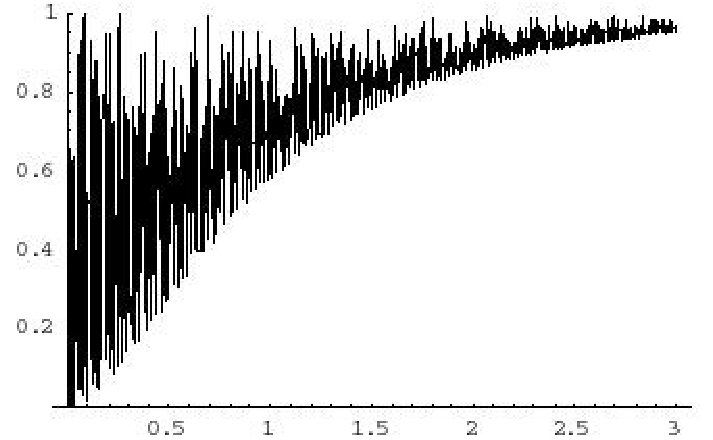}
}
%\vspace{5cm}       % Give the correct figure height in cm
\caption{$\gamma_{1,1}=1 KHz$, $\gamma_{2,2}=1 KHz$, $\gamma_{1,2}=\gamma_{2,1}=1 KHz$ , $\epsilon=47\pi KHz$
, t=10=1ms}
\label{fig:3}       % Give a unique label
\end{figure}

\begin{figure}
\resizebox{0.5\textwidth}{!}{%
  \includegraphics{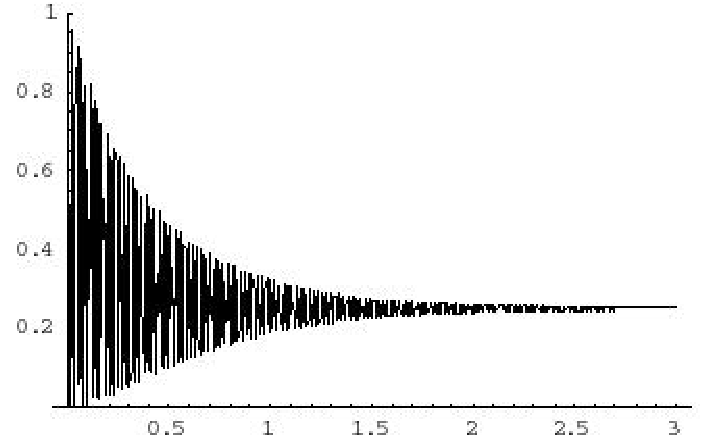}
}
%\vspace{5cm}       % Give the correct figure height in cm
\caption{$\gamma_{1,1}=1 KHz$, $\gamma_{2,2}=1 KHz$, $\gamma_{1,2}=\gamma_{2,1}=1 KHz$ , $\epsilon=47\pi KHz$
, t=10=1ms}
\label{fig:4}       % Give a unique label
\end{figure}
This is the main result of the paper which means that a successful
measurement, performed at large enough time instants, generates un
uncorrelated state of the two subsystems, bipartite system and
reservoir, leaving atom and cavity in their maximally antisymmetric
entangled state.

This ideal result has to be corrected by the introduction in the
microscopical model of a second bath of interaction able to take
account of the cavity leakage of energy because of the imperfect
mirrors. In terms of the hamiltonian operator this means:
\begin{eqnarray}
 H&=&\hbar\frac{\omega_0}{2}S_z+\hbar \omega_0 \alpha^\dag
 \alpha+\hbar(\epsilon\alpha S_+ +\epsilon^\ast\alpha^\dag S_-) \\ \nonumber &+&
\hbar\sum_{\vec{k},\lambda}\omega_{\vec{k}}
b^\dag_{\lambda}(\vec{k})b_{\lambda}(\vec{k})
  + \hbar\sum_{\vec{k},\lambda}\omega_{\vec{k}}
c^\dag_{\lambda}(\vec{k})c_{\lambda}(\vec{k}) \\ \nonumber &+&\sum_{\vec{k},\lambda}[g^\ast_{\vec{k},\lambda}
b_{\lambda}(\vec{k}) +g_{\vec{k},\lambda}
  b^{\dag}_{\lambda}(\vec{k})](S_+ +S_-)\\\nonumber&+&\sum_{\vec{k},\lambda}[s^\ast_{\vec{k},\lambda}
  b_{\lambda}(\vec{k})+s_{\vec{k},\lambda}
  b^{\dag}_{\lambda}(\vec{k})](\alpha^\dag +\alpha) \\ \nonumber &+&\sum_{\vec{k},\lambda}[r^\ast_{\vec{k},\lambda}
  c_{\lambda}(\vec{k})+r_{\vec{k},\lambda}
  c^{\dag}_{\lambda}(\vec{k})](\alpha^\dag +\alpha).
\end{eqnarray}

If we make the position

\begin{eqnarray}
 H_S=\hbar\frac{\omega_0}{2}S_z+\hbar \omega_0 \alpha^\dag
 \alpha+\hbar(\epsilon\alpha S_+ +\epsilon^\ast\alpha^\dag S_-),
\end{eqnarray}

\begin{eqnarray}
 H_B=\hbar\sum_{\vec{k},\lambda}\omega_{\vec{k}}
b^\dag_{\lambda}(\vec{k})b_{\lambda}(\vec{k})+\hbar\sum_{\vec{k},\lambda}\omega_{\vec{k}}
c^\dag_{\lambda}(\vec{k})c_{\lambda}(\vec{k}),
\end{eqnarray}

\begin{eqnarray}
 H_I=\sum_{\alpha=1}^2 A^{\alpha}B^{\alpha},
\end{eqnarray}
where
\begin{eqnarray}
B^1=\sum_{\vec{k},\lambda}[g^\ast_{\vec{k},\lambda}
b_{\lambda}(\vec{k}) +g_{\vec{k},\lambda}
  b{^\dag}_{\lambda}(\vec{k})],
\end{eqnarray}

\begin{eqnarray}
B^2&=&\sum_{\vec{k},\lambda}[s^\ast_{\vec{k},\lambda}
b_{\lambda}(\vec{k})+s_{\vec{k},\lambda}
  b{^\dag}_{\lambda}(\vec{k})] \\ \nonumber &+& \sum_{\vec{k},\lambda}[r^\ast_{\vec{k},\lambda}
c_{\lambda}(\vec{k})+r_{\vec{k},\lambda}
  c{^\dag}_{\lambda}(\vec{k})]
\end{eqnarray}
\begin{eqnarray}
A^1=(S_+ +S_-),
\end{eqnarray}
\begin{eqnarray}
A^2=(\alpha^\dag +\alpha)
\end{eqnarray}

we can describe the reduced dynamics of the bipartite system at
$T=0$ by a standard Master Equation and we can solve it in the same
way of the previous case. The changes in the microscopical model do
not introduce variation in the formal solution. Instead, the
presence of two different channel of dissipation modifies the
dynamical behavior. Actually, the system has now the possibility to
loose energy in environments that do not speak each other. This
means that, when the time is much longer than the sum of the single
emission time, the coherence induced by the common bath during the
transient will go to zero in the long time domain. Despite this
fact, the dechoerence time can be made as long as we need to
implement the required quantum protocol. Actually, named $k$ the
cavity decay rate, if $k$ is much greater than
$\gamma_{1,1}=\gamma_{2,2}$, then the storage of energy can be
maximized for a time sufficient to realize the quantum protocol.
\section{Conclusion}
\label{sec:6}
In this paper we have considered the interaction of a Jaynes and
Cummings system with the electromagnetic field (and with another
phenomenological zero temperature bath) in its vacuum state and,
solving the dynamical problem, we have analyzed the amount of
entanglement induced in the bipartite system by the common
electromagnetic reservoir. This has allowed us to quantitatively
characterize the regime under which field-induced cooperative
effects are not vanished by dissipation. Once the Decoherence Free
Regime is reached, transient entanglement tends to become stationary
and, therefore, usable for quantum gate implementation.

The asymptotic solution of the dynamical problem appears to be a
statistical mixture of a maximally entangled state and the ground
state of the open system, the probability to obtain one or the
others being the same. In the whole temporal domain the found
solution tell us that the state of the system is a statistical
mixture of the free energy system eigenoperators. This fact is
general enough \cite{nud} and it is consistent with the existence of
a photon detector device because the act of measurement introduces a
stochastic variable respect to which we can only predict the
probability to have one or another of the possible alternative
measures \cite{nud}. These probabilities can be regarded as the
weight of the possible alternative generalized trajectories. With
this approach the dynamics has to be depicted as a statistical
mixture of this alternative generalized trajectories. Moreover the
found trajectories evolve in time in a deterministic way: for
example the trajectory relative to the initially excited system
state is a shifted free evolution characterized by complex
frequencies that means an exponential decay free evolution. This
statement may give the sensation that every system has to decay in
its ground state because of the observed dynamics. It is in general
not true. Actually, if the system is multipartite as ours, it is
possible that it admits excited and entangled equilibrium
Decoherence Free Subspace (DFS) \cite{Lidar}(so as it happens in
some highly symmetric models), constituted by states on which the
action of $H_I$ is identically zero. If the system, during
evolution, passes through one of these states, the successive
dynamics will be decoupled from the environment evolution. An
equilibrium condition is reached in which entanglement is embedded
in the system.
%
% BibTeX users please use
% \bibliographystyle{}
% \bibliography{}

\end{document}